\documentclass[aps,prl,twocolumn, superscriptaddress,showpacs]{revtex4}

\usepackage[T1]{fontenc}
\usepackage[latin1]{inputenc}
\usepackage[dvips]{graphicx,color}
\usepackage{rotating}
\usepackage{bm}        
\usepackage{amssymb}   
\usepackage{amsmath} 
\usepackage{braket}
\def\pra#1#2#3{Phys.~Rev.~A~{\bf #1},\ #2\ (#3)}
\def\prl#1#2#3{Phys.~Rev.~Lett.~{\bf #1},\ #2\ (#3)}

\def\jpb#1#2#3{J. Phys. B {\bf #1},\ #2\ (#3)}
\def\rmp#1#2#3{Rev.~Mod.~Phys.~{\bf #1},\ #2\ (#3)}

\def\k1{k_1}
\def\k2{k_2}
\def\q1{q_1}
\def\q2{q_2}

\def\({\left (}
\def\){\right )}
\def\[{\left [}
\def\]{\right ]}

\newcommand{\beq}{\begin{equation}}
\newcommand{\eeq}{\end{equation}}

\newcommand{\threejm}[6]{ \left(\begin{array}{ccc} #1 & #3 & #5\\
                                              #2 & #4 & #6
                                \end{array}
                          \right)}

\begin{document}
\date{\today}
\title{Spin-orbit interactions and quantum spin dynamics in cold ion-atom collisions}

\author{Timur V. Tscherbul}
\affiliation{Chemical Physics Theory Group, Department of Chemistry, and Center for Quantum Information and Quantum Control, University of Toronto, Toronto, Ontario, M5S 3H6, Canada}
\affiliation{Department of Physics, University of Nevada, Reno, NV, 89557, USA}\email[]{ttscherbul@unr.edu}
\author{Paul Brumer}
\affiliation{Chemical Physics Theory Group, Department of Chemistry, and Center for Quantum Information and Quantum Control, University of Toronto, Toronto, Ontario, M5S 3H6, Canada}
\author{Alexei A. Buchachenko}
\affiliation{Skolkovo Institute of Science and Technology, 100 Novaya str., Skolkovo, Moscow Region 143025, Russia}
\affiliation{Institute of Problems of Chemical Physics RAS, Chernogolovka, Moscow Region 142432, Russia}

\begin{abstract}
We present accurate {\it ab initio} and quantum scattering calculations on a prototypical hybrid ion-atom system Yb$^+$-Rb, recently suggested as a promising candidate for the experimental study of open quantum systems, quantum information processing, and quantum simulation.  We identify the second-oder spin-orbit (SO) interaction as the dominant source of hyperfine relaxation in cold Yb$^+$-Rb collisions. Our results are in good agreement with recent experimental observations [L.~Ratschbacher {\it et al.}, \prl{110}{160402}{2013}] of hyperfine relaxation rates of trapped Yb$^+$  immersed in  an ultracold Rb gas. The calculated rates are 4 times smaller than predicted by the Langevin capture theory  and display a weak $T^{-0.3}$ temperature dependence,  indicating significant deviations from statistical behavior. Our analysis underscores the deleterious nature of the SO interaction and implies that  light ion-atom combinations  such as  Yb$^+$-Li should be used to minimize hyperfine relaxation and decoherence of trapped ions in ultracold atomic gases.

\end{abstract}

\maketitle

The exquisite controllability of trapped atomic and molecular ions is key to their use in emerging quantum technologies, including quantum information processing \cite{Ladd}, quantum simulation \cite{ManyBody1,ManyBody2}, and  precision measurement \cite{PrecisionMeasurement,PrecisionMeasurement2}. They also serve as ideal prototype systems for exploring quantum decoherence \cite{decoherence,Kohl13}, many-body physics \cite{ManyBody1,ManyBody2},  ultracold chemistry \cite{KohlNatPhys,JulienneNatPhys}, and astrochemistry \cite{IonsReactions}. In particular,  hybrid ion-atom systems consisting of trapped ions immersed in an ultracold gas of neutral atoms display a remarkably rich dynamical behavior \cite{Vuletic09,Hudson,KohlNature,KohlNatPhys,JulienneNatPhys,Kohl10,Kohl13,Dulieu,Hudson2, Denschlag}. Several experimental groups have observed thermalization, inelastic relaxation, chemical reactions, and three-body recombination to occur in ultracold collisions of Yb$^+$+Yb \cite{Vuletic09},  Yb$^{+}$+Ca \cite{Hudson}, Yb$^{+}$+Rb \cite{KohlNature,KohlNatPhys,Kohl10,Kohl13}, Ca$^{+}$+Rb \cite{Dulieu}, Ba$^+$+Ca \cite{Hudson2},  and Ba$^{+}$+Rb \cite{Denschlag}. A major goal of these experiments is to achieve sympathetic cooling of the ion by using ultracold atoms as a cooling medium \cite{Vuletic12,Hudson3}.

In a recent experimental realization of such a hybrid ion-atom system, K\"ohl and co-workers   immersed a single trapped Yb$^+$ ion in an ultracold cloud of spin-polarized Rb atoms \cite{KohlNature,KohlNatPhys,Kohl10,Kohl13}. While momentum-changing collisions with ultracold Rb atoms led to efficient cooling of the heavy ion, K\"ohl {\it et al.} observed unexplainably rapid spin  relaxation and decoherence  \cite{Kohl13}. As both the ion and the atom were initially prepared in their fully spin-polarized internal states, these surprising results suggest the presence of an efficient spin-changing mechanism, which destroys  spin coherence and prevents quantum information storage in the ion's internal degrees of freedom. The observation of large relaxation and decoherence rates \cite{Kohl13} casts doubt on the suitability of hybrid ion-atom systems for quantum information and precision measurement applications. 
It remains unclear, however, whether the observed relaxation and coherence-destroying mechanisms  \cite{Kohl13} are universal or specific to the Yb$^+$-Rb system.

Accurate quantum scattering calculations based on  {\it ab initio} interaction potentials reported for several ion-atom systems \cite{Julienne09,Julienne11,Koch} provide valuable insight into the mechanisms of  cold ion-atom collisions and enable the development of multichannel quantum defect models \cite{BoGao} and semiclassical approximations \cite{RobinAlex03,Peng09}. Useful as they are, these calculations  do not take into account spin-nonconserving  interactions, such as the  magnetic dipole and second-order SO interactions, which play an important role in collisions of  highly magnetic \cite{Pfau,Zoran} and heavy \cite{Julienne96,CsCs} neutral atoms, causing  rapid two-body losses similar to those observed experimentally for Yb$^+$-Rb \cite{Kohl13}. The long-range polarization interaction leads to a large number of partial wave contributions to ion-atom  scattering even at the lowest collision energies attainable in current experiments ($\sim$100~mK). 
The spin-nonconserving interactions break the rotational symmetry of the scattering problem and couple the partial wave  states with the spin states  of the ion-atom collision complex \cite{CsCs,prl10}, dramatically increasing the computational complexity of the calculations.   As a result, the effects of these interactions on ultracold ion-atom collisions remain completely unexplored.



Here, we report accurate quantum scattering calculations on the prototypical heavy ion-atom collision system Yb$^+$-Rb  studied in recent experiments \cite{Kohl13}. We solve the ion-atom quantum scattering problem exactly using state-of-the-art  {\it ab initio} molecular potentials and SO coupling matrix elements of the (YbRb)$^+$ complex \cite{Alexei13}.  We obtain quantitative agreement with the measured relaxation rates for $^{171}$Yb$^+$-Rb collisions and identify the second-order SO interaction as the dominant source of rapid collisional spin relaxation. These results demonstrate that modern {\it ab initio} and quantum scattering calculations can predict the  collisional properties of hybrid ion-atom systems with quantitative accuracy.  They strongly  suggest that  light  ion-atom combinations such as Yb$^+$-Li, where the second-order SO interaction is much weaker,  should be used in  experimental applications that require long spin relaxation and coherence times, such as quantum information processing, quantum  simulation, and precision measurement.


\begin{figure}[t]
	\centering
	\includegraphics[width=0.43\textwidth, trim = 0 0 0 0]{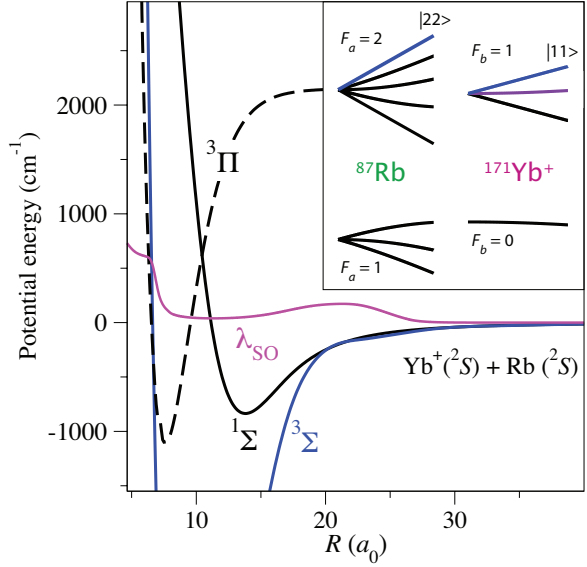}
	\renewcommand{\figurename}{Fig.}
	\caption{Scalar-relativistic interaction potentials for Yb$^+$-Rb of $^1\Sigma$ and $^3\Sigma$ (full lines) and $^3\Pi$ (dashed line) symmetries and the second-oder SO coupling [blue (grey) line] as a function of $R$. The inset shows the magnetic field dependence of the lowest hyperfine energy levels of $^{171}$Yb$^+$ and $^{87}$Rb (Zeeman splittings are exaggerated for clarity). The initial states chosen for scattering calculations are highlighted in color.}\label{fig:interactions}
\end{figure}

The quantum spin dynamics in Yb$^+$-Rb collisions is described by the  Hamiltonian \cite{Julienne09,Julienne11,Julienne96}
\begin{multline}\label{H}
\hat{H} = -\frac{1}{2\mu R}\frac{\partial^2}{\partial R^2}R + \frac{ \hat{\mathbf{L}}^2 }{2\mu R^2} + \hat{H}_a + \hat{H}_b +\hat{V}(R) \\ + \sqrt{\frac{24\pi}{5}} \left[ -\frac{\alpha^2}{R^3} + \lambda_\text{SO}(R)\right] \sum_{q} Y^\star_{2q}(\hat{R})[\hat{\mathbf{S}}_{a}\otimes\hat{\mathbf{S}_b}]^{(2)}_q,
\end{multline}
where $\hat{\mathbf{S}}_i$ are the electron spins of Rb ($i=a$) and $^{171}$Yb$^+$ ($i=b$), $\mu$ is the reduced mass of the Yb$^+$-Rb  collision complex,  $R$ is the internuclear separation, $\hat{\mathbf{L}}$ is the orbital angular momentum of the complex, and $\hat{R}$ describes the orientation of the complex in the laboratory frame with the quantization axis $z$ defined by the magnetic field vector $\mathbf{B}$. The asymptotic Hamiltonian of atom $i$ in Eq. (\ref{H}) is $\hat{H}_i = \gamma_{i}\hat{\mathbf{I}}_i\cdot \hat{\mathbf{S}}_i + 2\mu \mathbf{B} \cdot\hat{\mathbf{S}}_{i}$  \cite{Julienne09,Julienne11,Julienne96}, where $\hat{\mathbf{I}}_i$ is the atom's nuclear spin,  $\hat{V}(R)=\sum_{S,M_S}V_{SM_S}(R)|SM_S\rangle \langle SM_S|$ is the interaction potential which depends on the total spin $\hat{\mathbf{S}}=\hat{\mathbf{S}}_a+\hat{\mathbf{S}}_b$ of the collision complex and its projection $M_S$ on the $B$-field axis, and $\alpha$ is the fine structure constant. Figure~1 shows the relevant Yb$^+$-Rb potentials of $^1\Sigma^+$ and $^3\Sigma^+$  symmetry (correlating to the Yb$^+(^2S)$-Rb($^2S$) limit) obtained from high-level {\it ab initio} calculations \cite{Alexei13}.  These potentials are accurate enough   to yield collision-induced charge transfer (CCT) rates in quantitative agreement with experiment \cite{Alexei13,Lane1,Lane2,Kohl10}. The collisional processes of interest here occur on timescales much shorter  than CCT ($\ll$10 s), so we neglect the weak coupling to the ground  $X^1\Sigma^+$ state of Yb-Rb$^+$ \cite{Kohl10,Alexei13}.


An essential new aspect of this work, as compared to the previous theoretical studies of ion-atom collisions \cite{Julienne09,Julienne11,Koch}, is the presence of the $R$-dependent  SO interaction [the last term in Eq. (\ref{H})] between the $^3\Sigma^+$ and $^1\Sigma^+$ states, which does not conserve the total spin of the collision complex  and  causes inelastic transitions in  spin-polarized Yb$^+$-Rb collisions \cite{Julienne96,CsCs}.  
This interaction arises in the second order due to first-order couplings between the ground $\Sigma$ and excited $\Pi$ states.  As shown in Fig. 1, the $^3\Pi$ state, which correlates to the Yb$^+(^3P)$-Rb($^2S$) limit, crosses the potentials of both the $^1\Sigma$ and $^3\Sigma$ states at short range, leading to a resonant enhancement of the second-order SO coupling.  The magnitude of this coupling is proportional to the splitting between the relativistic  $^3\Sigma^+_0$ and $^3\Sigma^+_1$ components of the $^3\Sigma^+$ state, as described in the Supplemental Material \cite{SM}. 


We solve the ion-atom quantum scattering problem by expanding the stationary eigenfunctions of the Hamiltonian (\ref{H})  in direct-product basis functions $\phi_n(\hat{R})=|F_a m_{F_a}\rangle |F_b m_{F_b}\rangle |l m_l\rangle$, where $|F_i m_{F_i}\rangle$ are the atomic hyperfine states and $|lm_l\rangle$ are the eigenstates of $\hat{\bf{L}}^2$ and $\hat{L}_z$. The radial expansion coefficients $\mathcal{F}_n(R)$ satisfy a system of coupled-channel (CC)  equations
\begin{multline}\label{CC}
\left[ \frac{d^2}{dR^2} - \frac{l(l+1)}{R^2} +2\mu E \right] \mathcal{F}_n(R) \\ =  2\mu\sum_{n'} \langle \phi_n| \hat{V}(R) + \hat{H}_a +\hat{H}_b  + \hat{V}_\text{SO} |\phi_{n'} \rangle  \mathcal{F}_{n'}(R), 
\end{multline}
where $E$ is the total energy and $\hat{V}_\text{SO}$ stands for the second-order SO interaction [the last term in Eq. (\ref{H})].
 The matrix elements of the interaction potential and $\hat{H}_i$ in Eq. (\ref{CC}) are calculated as described elsewhere \cite{ZhiyingRoman07} whereas those of $\hat{V}_\text{SO} $ are derived in the Supplemental Material \cite{SM}.

The CC equations (\ref{CC}) are solved numerically at fixed total angular momentum projection $M=m_{F_a}+ m_{F_b}+m_\ell$ on a grid of $R\in [3,\, 3\times 10^4] a_0$ with a grid spacing of 0.01$a_0$. All basis states with $l\le 40$ are included in scattering calculations to produce converged results in the experimentally relevant range of collision energies of 40-240~mK \cite{Kohl13}, leading to a total of 1276 channels for $M=0$.  Scattering boundary conditions are applied after reaching the outer end of the integration grid to extract the scattering $S$-matrix elements, which are used to compute the total ($M$-summed) scattering cross sections and transition rates. 

\begin{figure}[t]
	\centering
	\includegraphics[width=0.47\textwidth, trim = 0 0 0 0]{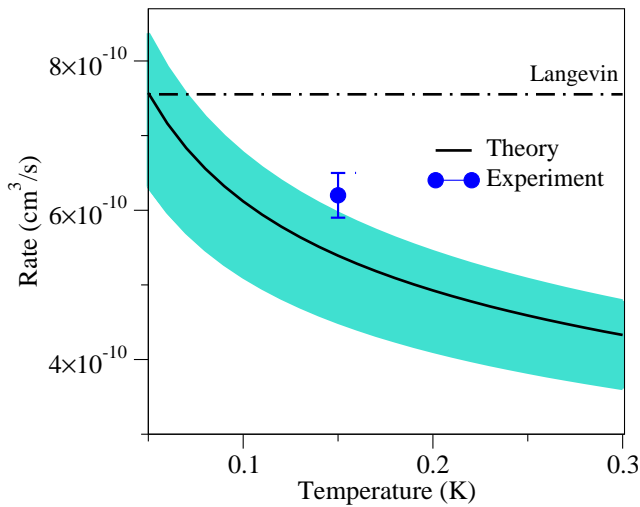}
	\renewcommand{\figurename}{Fig.}
	\caption{Inelastic rate constant for the $|11\rangle \to |00\rangle$ transition in Yb$^+$ induced by collision with Rb ($|2,2\rangle$). Full line -- theory (present work); circle with error bars -- experiment \cite{Kohl13}, dashed line -- Langevin  rate scaled down by a factor of 2.8 for clarity. The shaded area shows the uncertainty arising from the inaccuracies in {\it ab initio} interaction  potentials. }\label{fig:rates}
\end{figure}

Figure 2 shows the calculated inelastic rate constant for the $|1,1\rangle \to |0,0\rangle$  hyperfine transition in Yb$^+$ induced by collisions with spin-polarized Rb at $B=6$~G. First, we observe good agreement between the calculated and measured  rates \cite{Kohl13}.  
At $T=150$~mK, the calculated rate is 4 times smaller than the Langevin collision rate \cite{Gao11}, $k_L=2\pi \sqrt{C_4/\mu}=2.1\times 10^{-9}$~cm$^3$/s. Second, the exact quantum rate displays a weak $T^{-0.3}$ temperature dependence whereas the ion-atom Langevin rate is temperature-independent \cite{Kohl13}, indicating significant deviations from statistical behavior in ultracold Yb$^+$-Rb collisions.  Third, as both Yb$^+$ and Rb are fully spin-polarized prior to collision, the large magnitude of the inelastic rate can only be caused by a spin-nonconserving interaction. Test calculations show that omitting the magnetic dipole interaction from the Hamiltonian (\ref{H}) does not change the results, leading us to conclude that it is the second-order SO interaction  that is responsible for the rapid spin relaxation observed experimentally \cite{Kohl13}. 

To estimate the uncertainty of the theoretical results, we performed quantum scattering calculations with modified $^1\Sigma$ and $^3\Sigma$ potentials \cite{Alexei13} obtained by shifting the short-range parts of the potentials by a constant factor $\Delta R = \pm 0.02 a_0$. While this modification results in a large change of the $s$-wave scattering lengths,  the calculated inelastic rates at 40~mK vary only by  ${ }^{+10.6}_{-16.6}\%$ as shown in  Fig.~2.   This is because  Yb$^+$-Rb scattering at the collision energies of interest here ($\sim$20-100 mK) occurs in the multiple partial wave regime, where the resonance contributions due to individual partial waves are averaged out, and the rates are determined by the value of the SO coupling at the inner turning points of the interaction potentials. Figure~2 shows that the calculated upper limit of the inelastic rate agrees well with the measured value (see Table I), suggesting that left-shifted Yb$^+$-Rb potentials provide a better agreement with the measured inelastic rates.



Figure 3 shows normalized product state distributions 
\begin{equation}\label{dist}
 P( F_a, m_{F_a}; F_b, m_{F_b}) = k_{F_a, m_{F_a}, F_b m_{F_b}}/k_\text{inel}
\end{equation}
calculated from the inelastic rates $k_{F_a, m_{F_a}, F_b m_{F_b}}$, where  $k_\text{inel} = \sum_{F_a m_{F_a}, F_b m_{F_b}} k_{F_a, m_{F_a}, F_b m_{F_b}}$ is the total inelastic rate for a given  $ |2,2\rangle_a |F_b, m_{F_b}\rangle$  initial state.
 The inset of Fig.~3 shows the marginal distribution $P( F_b, m_{F_b})$ obtained by summing Eq. (\ref{dist}) over all final hyperfine states of Rb. While the populations of the hyperfine states $|1,0\rangle_b$ and $|0,0\rangle_b$ of Yb$^+$ are similar, collision-induced energy transfer  into the  $|1,-1\rangle_b$ state is about 50 times slower. This selection rule was assumed by K{\"o}hl and co-workers in their analysis of experimental data \cite{Kohl13} and our calculations  provide a rigorous justification of this assumption. The $|1,1\rangle_b \to |1,-1\rangle_b$ transition is suppressed in first order because the matrix element of the SO interaction \cite{SM} contains the 3-$j$ symbol $\threejm{F_b}{-m_{F_b}} {1}{m_{F_b}-m_{F_b}'}{F_b'}{m_{F_b}'}$, which vanishes identically for $F_b=F_b'=1$ and $m_{F_b}=1$, $m_{F_b}'=-1$. In contrast, the dominant transitions to the $|10\rangle_b$ and $ |00\rangle_b$ hyperfine states are allowed to first order.

Figure 3 shows that the product-state distributions (\ref{dist}) for the dominant transitions to the  final states $|1,0\rangle_b$ and $|0,0\rangle_b$ are peaked at the initial state $|2,2\rangle_a$ of Rb. While there is clear preference for the initial state $|2,2\rangle_a$ to remain unchanged in a  collision, hyperfine-changing transitions to the final states $|1,0\rangle_a$ and $|1,1\rangle_a$ also occur with significant probabilities ($\sim$15-25\%), which are weakly sensitive to the final hyperfine state of Yb$^+$.  The hyperfine distributions for the suppressed $|1,1\rangle_b \to |1,-1\rangle_b$ transition are, in contrast, peaked at the lowermost Rb state $|1,1\rangle_a$.

\begin{table}
\caption{Calculated and measured hyperfine relaxation rates (in units of $10^{-10}$ cm$^3$/s) for $F$- and $m_F$-changing transitions in Yb$^+$-Rb collisions at $B = 6$~G. All rates are computed from inelastic cross sections at 40 mK except for the $|11\rangle_b \to |00\rangle_b$ transition, for which the thermally averaged rate at $T = 150$~mK is given. } 
\centering
\begin{tabular}{ccc}
\hline\hline
Transition &  $\quad$ Theory  $\quad$ &   Experiment   \\
\hline
\hline
$|1,1\rangle_b \to |0,0\rangle_b$          &    $\quad$   $5.40^{+0.57}_{-0.90}$  $\quad$    &      6.2(0.3) \\
$|1,1\rangle_b \to |1,0\rangle_b$          &     $\quad$  3.33   $\quad$   &      3.4(0.6)   \\
$|1,1\rangle_b \to |1,-1\rangle_b$         &    $\quad$   0.19  $\quad$    &      0   \\
$|1,0\rangle_b \to |1,-1\rangle_b$         &    $\quad$   2.62  $\quad$    &      3.4(0.6)   \\
$|1,0\rangle_b \to |1,1\rangle_b$         &    $\quad$   3.62  $\quad$    &      5.1(0.6)   \\
\hline\hline
\end{tabular}
\end{table}


The experimental estimates of the Yb$^+$ temperature~$T$ were limited by the lack of insight into an important heating mechanism involving collisional de-excitation of the $|2,2\rangle_a$ hyperfine state of Rb \cite{Kohl13,LotharThesis}.  This temperature sets the collision energy with ultracold Rb atoms, and is given by $\epsilon T_\text{max}$, where $T_\text{max}=240$~mK and $\epsilon$ is the probability of Rb hyperfine state change in a Langevin collision \cite{Kohl13}.  To improve the experimental estimate of~$T$, we calculated  $\epsilon$ as a sum of transition probabilities to the $F_a'=1$ hyperfine manifold of Rb. For the relevant $|1,1\rangle_b \to |0,0\rangle_b$ transition in Yb$^+$, we calculate $\epsilon=\sum_{m_{F_a}} P (F_a = 1, m_{F_a}; F_b = 0,  m_{F_b} = 0) = 0.64$. An improved estimate of the ion temperature for comparison with theory  is thus $T=\epsilon \times$240 mK $\approx$ 150 mK.



Thus far we have focused on hyperfine transitions from a single initial state $|1,1\rangle_b$ in spin-polarized Yb$^+$-Rb collisions. Table I  compares the results of our scattering calculations for the other hyperfine transitions in Yb$^+$  with the measured values \cite{Kohl13}. 
We observe quantitative agreement between  experiment and theory for all the transitions except $|1,0\rangle_b \to |1,1\rangle_b$ and $|1,0\rangle_b \to |1,-1\rangle_b$, the rates of which  were not directly measured, but rather inferred from $^{174}$Yb$^+$ measurements \cite{Kohl13,LotharThesis} under several assumptions, including  (1) the relation $\gamma_\text{ex}/\gamma_{SR}= 0.5$ between the excitation and  relaxation rates based on the values observed for $^{174}$Yb$^+$ \cite{LotharThesis}; (2) the ratio $r$ of the  $|1,-1\rangle_b \to |10\rangle_b$ and $|10\rangle_b \to |11\rangle_b$ transition rates is equal to 1.5 (we find $r=1.4$), and (3) the transitions changing $m_F$ by 2 or more are strictly forbidden. Additionally, collision-induced hyperfine relaxation from other than the fully spin-polarized initial states of Yb$^+$ can proceed via the spin-exchange mechanism due to the different phase shifts associated with the singlet and the triplet potentials \cite{Li14,LiGao15}. As this mechanism is more sensitive to the uncertainties of the interaction potentials, we expect the calculated and experimentally derived $|1,0\rangle_b \to |1,1\rangle_b$ and $|1,0\rangle_b \to |1,-1\rangle_b$ transition rates to be more uncertain than the $|1,1\rangle_b \to |0,0\rangle_b$ transition rate. The hyperfine relaxation  rates are of the same order of magnitude (10$^{-10}$~cm$^3$/s) and they are not very sensitive to the initial state, which is  consistent with a strong spin-depenendent coupling mechanism between the internal states,  mediated by both  $S$-conserving spin-exchange and $S$-changing second-order SO interactions.


\begin{figure}[t]
	\centering
	\includegraphics[width=0.43\textwidth, trim = 0 0 0 0]{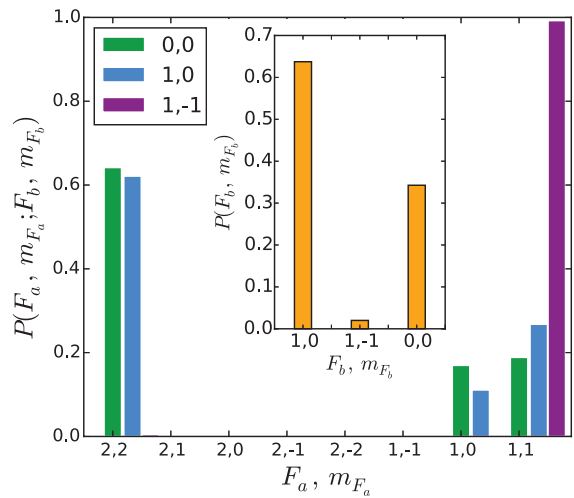}
	\renewcommand{\figurename}{Fig.}
	\caption{Normalized product state distributions $P(F_a,m_{F_a}; F_b, m_{F_b})$ (Eq. \ref{dist}) plotted for a collision energy of 40 mK and $B=6$ G. Adjacent bars correspond to the different  hyperfine states of Yb$^+$: $|00\rangle_b$ (left), $|10\rangle_b$ (middle), and $|1,-1\rangle_b$ (right).  The inset shows Yb$^+$ product state distributions  summed over all final hyperfine states of Rb.}\label{fig:dist}
\end{figure}

In summary, we have presented the first rigorous theoretical analysis of quantum spin dynamics in cold heavy ion-atom collisions. Unlike all previous theoretical models \cite{Julienne09, Julienne11,Koch}, our CC approach explicitly takes into account  spin-nonconserving  interactions, which play a critical role in collisions of heavy ions with coolant atoms. 
 Our calculations show that the lowermost $\Sigma$ states of heavy ion-atom complexes exhibit a short-range crossing with the states of $\Pi$ symmetry, giving rise to a strong second-order SO interaction (Fig.~1), which leads to rapid spin relaxation in cold ion-atom collisions. Our calculated spin relaxation rates are in good agreement with recent experiments  \cite{Kohl13} (Fig.~2 and Table~I). As the magnetic dipole interaction makes a negligible contribution to the overall Yb$^+$-Rb spin relaxation rate,  we conclude that heavy ion-atom collision systems exhibiting strong SO interactions (such as  Yb$^+$-Rb and Ba$^+$-Rb \cite{Krych11}) are unsuitable for quantum technological applications, which require long spin relaxation and coherence times. Rather, for these applications, it is advisable to use light coolant atoms such as Li or Na, where the SO interactions are weaker. Indeed, recent {\it ab initio} calculations \cite{Koch} suggest that the $^3\Pi$ electronic state of the Yb$^+$-Li complex does not cross the $\Sigma$ states, which indicates that the SO interaction in this system will be suppressed, leading one to expect favorably long spin relaxation and coherence times. Our quantum scattering approach can be used to investigate the dynamics of inelastic relaxation in both light and heavy  ion-atom  collision systems. It  can also be extended to study the mechanisms of collisional decoherence of atomic and molecular ions immersed in ultracold atomic buffer gases \cite{Hornberger}, for which the first experimental results have recently become available \cite{Kohl13}. Collisional decoherence is one of the most fundamental mechanisms responsible for the quantum-to-classical transition \cite{Schlosshauer}, and has so far been tested experimentally only at elevated temperatures \cite{Zeilinger}. Suppressing the decoherence mechanisms with external electromagnetic fields \cite{ChrisRoman} would be an important step toward quantum technological applications based on trapped ion-atom hybrid systems.

We thank Lothar Ratschbacher and Michael K\"ohl for the valuable discussions. This work was supported by the University of Nevada, Reno, NSERC of Canada and Russian FBR (projects 14-03-00422 and 14-33-50861).

\end{document}



\title{Supplemental Material for the manuscript \\ ``Spin-orbit interactions and quantum spin dynamics in cold ion-atom collisions''}

\author{Timur V. Tscherbul}
\affiliation{Chemical Physics Theory Group, Department of Chemistry, and Center for Quantum Information and Quantum Control, University of Toronto, Toronto, Ontario, M5S 3H6, Canada}
\affiliation{Department of Physics, University of Nevada, Reno, NV, 89557, USA}\email[]{Email: ttscherbul@unr.edu}
\author{Paul Brumer}
\affiliation{Chemical Physics Theory Group, Department of Chemistry, and Center for Quantum Information and Quantum Control, University of Toronto, Toronto, Ontario, M5S 3H6, Canada}
\author{Alexei A. Buchachenko}
\affiliation{Skolkovo Institute of Science and Technology, 100 Novaya str., Skolkovo, Moscow Region 143025, Russia}
\affiliation{Institute of Problems of Chemical Physics RAS, Chernogolovka, Moscow Region 142432, Russia}

\maketitle


In this Supplemental Material we describe the evaluation of the matrix element of the second-order spin-orbit (SO) interaction based on  the  {\it ab initio} (YbRb)$^+$ interaction potentials computed in Ref. \cite{Alexei13}. In Sec. I we derive the relationship $\lambda_\text{SO}(R) = -\frac{2}{3}\Delta_\text{SO}$, where $\lambda_\text{SO}(R)$ is the matrix element of the SO coupling in Eq. (1) of the main text and $\Delta_\text{SO}$ is the splitting between the relativistic electronic states of $a0^-$ and $a1$ symmetries \cite{Alexei13}. In Sec. II we evaluate the matrix elements of the second-order SO interaction in the  scattering basis defined in the main text. These matrix elements are necessary to parametrize the coupled-channel equations [Eq. (2) of the main text].

\section{Second-order SO interaction from {\it ab initio} calculations}  


To obtain the phenomenological second-order SO interaction parameter $\lso(R)$, consider the $\Sigma=0$ and $\Sigma = \pm 1$ components of the $a^3\Sigma$ scalar-relativistic state of Yb$^+$--Rb \cite{Alexei13}, for which $\Omega = \Sigma$. In the absence of the magnetic dipole-dipole interaction and the second-order SO interaction,  these components are exactly degenerate. The perturbation Hamiltonian responsible for the splitting is \cite{Julienne08}
\begin{equation}\label{H}
\hat{H}_\text{dip}^\text{eff} = \left( -\frac{\alpha^2}{R^3} + \lso(R) \right) [3\hat{S}_{a_z} \hat{S}_{b_z} - \hat{\bf{S}}_a \cdot \hat{\bf{S}}_b],
\end{equation}
where $\alpha$ is the fine-structure constant, $R$ is the distance between the atoms, $\hat{S}_i$ is the electron spin of atom $i=a,b$ and $\hat{S}_{i_z}$ is the projection of $\hat{S}_i$ on the {\it internuclear (molecule-fixed)} axis. The second-order SO interaction is parametrized by an $R$-dependent parameter $\lso(R)$. The term proportional to $1/R^3$ describes the contribution due to the magnetic dipole-dipole interaction.

In general, both the second-order SO interaction and the magnetic dipole-dipole interaction enter the prefactor in Eq.~(\ref{H}) and contribute to the splitting $\Delta_\text{SO}= V_{\Sigma=0}(R) - V_{\Sigma=1}(R)$. However, because the magnetic  dipole-dipole interaction is not  included in the {\it ab initio} calculations,  the splitting between the potential energy curves of $a0^-$ and $a1$ symmetries is due to the second-order SO interaction alone, and Eq. (\ref{H}) can be written as
\begin{equation}\label{SO}
\hat{H}_\text{SO} = \lso(R)  [3\hat{S}_{a_z} \hat{S}_{b_z} - \hat{\bf{S}}_a \cdot \hat{\bf{S}}_b] = \lso(R) \sqrt{6} [\hat{\bf{S}}_a\otimes \hat{\bf{S}}_b]^{(2)}_0,
\end{equation}
where we expressed the product of two $\hat{S}_z$ operators in tensor form (Ref. \cite{Zare}, Eq. 5.44) neglecting the constant   terms, which only cause an overall energy shift.

We take the angular momentum basis functions for the $\Sigma = 0$ and $\Sigma =\pm1$ components of the triplet state to be
\begin{equation}
|(S_a S_b) S\Sigma\rangle = \sum_{\Sigma_a,\,\Sigma_b} \cgc{S_a}{S_b}{S}{\Sigma_a}{\Sigma_b}{\Sigma} |S_a\Sigma_a\rangle |S_b\Sigma_b\rangle,
\end{equation}
where $[:::]$ are the Clebsh-Gordan coefficients.

Now we seek to evaluate the splitting between the expectation values of the SO Hamiltonian (\ref{SO}) over the basis functions with the same total spin $S=1$ but different $\Sigma=0, \, \pm1$
\begin{equation}\label{delta}
\Delta_\text{SO} = \langle (S_a S_b) 10 |\hat{H}_\text{SO}|  (S_a S_b) 10 \rangle  -   \langle (S_a S_b)11 |\hat{H}_\text{SO}|  (S_a S_b) 11\rangle,
\end{equation}
where it is understood that $S_A = S_B = 1/2$. To evaluate the matrix elements entering this expression, we first use the Wigner-Eckart theorem to show that
\begin{equation}\label{WET}
\langle (S_a S_b) S\Sigma |   [\hat{\bf{S}}_a\otimes \hat{\bf{S}}_b]^{(2)}_0 |  (S_a S_b) S\Sigma \rangle  
= \ (-1)^{S-\Sigma}  \threejm{S}{-\Sigma}{2}{0}{S}{\Sigma}  
\langle (S_a S_b) S ||   [\hat{\bf{S}}_a\otimes \hat{\bf{S}}_b]^{(2)} ||  (S_a S_b) S \rangle. 
 \end{equation}
The reduced matrix element can be evaluated using Eq. (5.68) of Ref. \cite{Zare} to yield the result
\begin{multline}\label{ME}
 \langle (S_a S_b) S\Sigma |   \hat{H}_\text{SO} |  (S_a S_b) S\Sigma \rangle  
= \sqrt{30} \lso(R) (-1)^{S-\Sigma}  \threejm{S}{-\Sigma}{2}{0}{S}{\Sigma} (2S+1) \ninej{S_a}{S_a}{1}{S_b}{S_b}{1}{S}{S}{2} \\ \times [(2S_a+1)S_a(S_a+1)]^{1/2}
[(2S_b+1)S_b(S_b+1)]^{1/2}.
 \end{multline}
Substituting the values of $\Sigma = 0$ and $1$, $S = 1$, and $S_a = S_b = 1/2$, and using Eq. (\ref{delta}), we find
\begin{equation}\label{delta2}
\langle 10 |   \hat{H}_\text{SO} |  10 \rangle   = -\lso(R); \,\, \langle 11 |   \hat{H}_\text{SO} |  11 \rangle   = \frac{1}{2}\lso(R)
\end{equation}
and hence
\begin{equation}\label{result}
\lso = -\frac{2}{3}\Delta_\text{SO}.
\end{equation}
We note that this result is consistent with Eq. (13d) of Ref. \cite{Julienne96}, where the SO splitting is given by second-order perturbation theory 
(neglecting the small contribution due to the $^1\Pi$ state) 
\begin{equation}\label{J96}
P_1 = - \frac{ |\langle ^3\Sigma_{u,\Omega=1} |H_\text{SO}| ^3\Pi_{u,\Omega=1} \rangle|^2}{V(^3\Pi_u, R) - V(^3\Sigma_u,R)}
\end{equation}
 and the desired  SO parameter (analogous to our $\lso(R)$) is given by Eq. (14) as $V^\text{SO}_{\Omega=0}(R)= \frac{2}{3} P_1$.
 
{Here, we evaluate $\Delta_\mathrm{SO}$ as the difference between the $a0^-$ and $a1$ potential energy curves calculated by diagonalizing the {\it ab initio} SO Hamiltonian matrix in the basis spanning
  all non-relativistic states of the (YbRb)$^+$ complex correlating to the Yb($^1$S) + Rb$^+$($^1$S), Yb$^+$($^2$S) + Rb($^2$S) and Yb($^3$P$^\circ$) + Rb$^+$($^1$S) dissociation limits \cite{Alexei13}. The approach thus implicitly goes beyond the second-order approximation assumed in the above derivation.}  
 
 \section{Matrix elements of the second-order SO interaction}
 
In this section, we evaluate the matrix elements  of the second-order SO interaction in the direct-product scattering basis  $|\phi_n\rangle =|F_a m_{F_a}\rangle |F_b m_{F_b}\rangle |l m_l\rangle$ 
\begin{multline}\label{SOmel}
\langle \phi_n| \hat{V}_\text{SO} |\phi_n'\rangle = \sqrt{\frac{24\pi}{5}} \lambda_\text{SO}(R) \sum_q  (-1)^q  \langle l m_l| Y_{2,-q}(\hat{R})  | l' m_l'\rangle  \\ \times
\langle (I_a S_a) F_a m_{F_a} | \langle (I_b S_b) F_b m_{F_b} |    [\hat{\mathbf{S}}_a \otimes \hat{\mathbf{S}}_b]^{(2)}_q |  (I_a S_a) F_a' m_{F_a}' \rangle |  (I_b S_b) F_b' m_{F_b}' \rangle, 
\end{multline}
where $F_i$ are the total angular momenta of the Rb atom ($i=a$) and the Yb$^+$ ion ($i=b$), $l$ is the orbital angular momentum for the collision, and $m_{F_i}$ and $m_l$ are the projections of $F_i$ and $l$ on the magnetic field vector (which sets a space-fixed quantization axis).
To simplify the notation, we will heretofore omit the indices referring to the electron and nuclear spins of the atom and the ion ($I_a, S_a$ and $I_b, S_b$) except when necessary to avoid confusion, since these angular momenta remain unchanged in a collision.

By the definition of the tensor product, the second line in Eq. (\ref{SOmel}) can be factorized as
\begin{equation}\label{tensorproduct}
\sum_{q_a, q_b} (-1)^q \sqrt{5} \threejm{1}{q_a}{1}{q_b}{2}{-q} \langle F_a m_{F_a} | (\hat{S}^{(1)}_a)_{q_a} | F_a' m_{F_a}'\rangle
\langle F_b m_{F_b} | (\hat{S}^{(1)}_b)_{q_b} | F_b' m_{F_b}'\rangle,
\end{equation}
where $(\hat{S}^{(1)}_a)_{q_a}$ and $(\hat{S}^{(1)}_a)_{q_a}$ are the spherical tensor operators composing the tensor product.
The matrix elements of these operators over the hyperfine basis functions $|F_a m_{F_a}\rangle$ can be obtained  as follows.  First, we use  the Wigner-Eckart theorem \cite{Zare} to write
\begin{equation}\label{wet}
 \langle (I_a S_a) F_a m_{F_a} | (\hat{S}^{(1)}_a)_{q_a} | (I_a S_a) F_a' m_{F_a}'\rangle = (-1)^{F_a- m_{F_a}} \threejm{F_a}{-m_{F_a}}{1}{q_a}{{F_a}'}{m_{F_a}'}      \langle (I_a S_a) F_a  || \hat{S}^{(1)}_a || (I_a S_a) F_a' \rangle 
 \end{equation}
and similarly for the matrix element of $(\hat{S}^{(1)}_b)_{q_b}$. The double-bar matrix element on the right-hand side is \cite{Zare}
\begin{equation}\label{doublebar}
    \langle (I_a S_a) F_a  || (\hat{S}^{(1)}_a) || (I_a S_a) F_a' \rangle = (-1)^{I_a+S_a+F_a+1}[(2F_a+1)(2F_a'+1)]^{1/2} \sixj{S_a}{F_a'}{F_a}{S_a}{I_a}{1} \langle S_a || \hat{S}^{(1)}_a || S_a\rangle,
 \end{equation}
where $\{...\}$ is a 6-$j$ symbol and $\langle S_a || \hat{S}^{(1)}_a || S_a\rangle = [(2S_a+1)S_a(S_a+1)]^{1/2}$ \cite{Zare}. We thus get 
 \begin{multline}\label{wet}
 \langle (I_a S_a) F_a m_{F_a} | (\hat{S}^{(1)}_a)_{q_a} | (I_a S_a) F_a' m_{F_a}'\rangle = (-1)^{I_a + S_a+2F_a- m_{F_a}+1}[(2F_a+1)(2F_a'+1)]^{1/2}  [(2S_a+1)S_a(S_a+1)]^{1/2} \\ \times \threejm{F_a}{-m_{F_a}}{1}{q_a}{F_a'}{m_{F_a}'}   \sixj{S_a}{F_a'}{F_a}{S_a}{I_a}{1}.  \end{multline}
Similar expressions follow for the matrix element of $(\hat{S}^{(1)}_b)_{q_b}$.

Using the standard integral over three spherical harmonics \cite{Zare}
 \begin{equation}\label{wet}
\langle lm_l | Y_{2,-q}(\hat{R}) | l' m_l'\rangle = (-1)^{m_l} \left[\frac{(2l+1)5(2l'+1)}{4\pi}\right]^{1/2} \threejm{l}{0}{2}{0}{l'}{0} \threejm{l}{-m_l}{2}{-q}{l'}{m_l'}
 \end{equation}
and collecting all the terms, we obtain the matrix element of the second-order SO interaction as
 \begin{multline}\label{SOmelFinal}
\langle \phi_n| \hat{V}_\text{SO} |\phi_n'\rangle = \lambda_\text{SO}(R) (-1)^{I_a+S_a +I_b +S_b +2F_a +2F_b - m_{F_a} - m_{F_b} + m_l}  \sqrt{30} [(2l+1)(2l'+1)]^{1/2}  [(2S_a+1)S_a(S_a+1)]^{1/2} \\ \times [(2S_b+1)S_b(S_b+1)]^{1/2} [(2F_a+1)(2F_a'+1)]^{1/2}  [(2F_b+1)(2F_b'+1)]^{1/2} \sixj{S_a}{F_a'}{F_a}{S_a}{I_a}{1}  \sixj{S_b}{F_b'}{F_b}{S_b}{I_b}{1}  \\ \times
 \sum_q \threejm{l}{0}{2}{0}{l'}{0}  \threejm{l}{-m_l}{2}{-q}{l'}{m_l'} \sum_{q_a, \, q_b} \threejm{1}{q_a}{1}{q_b}{2}{-q} \threejm{F_a}{-m_{F_a}}{1}{q_a}{F_a'}{m_{F_a}'}\threejm{F_b}{-m_{F_b}}{1}{q_b}{F_b'}{m_{F_b}'}.
\end{multline}
It follows from the symmetry properties of the 3-$j$ symbols that $q_a = m_{F_a} - m_{F_a}'$, $q_b = m_{F_b} - m_{F_b}'$, and $q = q_a + q_b = m_l - m_l'$. These equalities reflect the fact that the basis functions with different the total angular momentum projections $M=m_{F_a} + m_{F_b} + m_l$ are not coupled by the SO interaction (or any other interaction in the Hamiltonian, see Eq. (1) of the main text). Thus, as expected, $M$ remains  a good quantum number for ion-atom collisions in a magnetic field.